# An Edge-based Graph Transformer Network for Anti-Cancer Drug Synergy Prediction

Jie Hu[a]

[a] *School of Computing/Software, University of South China, Hunan 421001, China;*

**Abstract**

Drug combination therapy is a powerful solution for the treatment of complex disease such as cancers due to its capability of therapeutic efficacy and reducing side effects. Nevertheless, it is very difficult to screen all drug combinations by experiments since the vast number of possible combinations. Currently, computational methods, especially graph neural networks and transformer, have been developed to discover the prioritization of drug combinations and shown promising potentials. Despite great achievements have been obtained by existing computational models, they all neglected high-order semantic information of drugs and the importance of the chemical bond features, which contained rich information and is represented by edge of graphs in drug predictions. In this work, we present a novel model named EGTSyn (*Edge-based Graph Transformer network for drug Synergy prediction*) for anti-cancer drug synergistic effect prediction. We design an EGNN (*edge-based graph neural network*) module and a GTDblock (*Graph Transformer for Drugs block*). EGNN is employed to capture the global structure information of the chemicals as well as the importance of chemical bonds that has been neglected by most of the previous studies. GTDblock combines the EGNN module with a Transformer-architecture encoder to extract high-order semantic information of drugs. The proposed EGTSyn is compared with the baseline methods including state-of-the-art deep learning based models and traditional machine learning based models. Experimental results show that EGTSyn has superior performance over these competitive baseline methods. Meanwhile, the leave-one-out experiments are conducted to further verify the performance of EGTSyn and the results are all quite positive. In addition, we test our trained model on the unseen drug pairs and the results verify the good generalization performance of EGTSyn. As proved by these experiments, the proposed EGTSyn has an excellent performance for anti-cancer drug-combinations synergistic effect prediction and is quite promising for further wet-lab research.

## Methodology

### *1. Data collections*

The benchmark drug synergistic combination data were obtained from a large-scale oncology screening data set which was released by recent studies [1, 2]. Conditioned by cancer cell line expression data, 12415 drug pairs were selected to build the benchmark dataset and it covers 31 human cancer cell lines and 36 anti-cancer drugs. The degree of synergy indication is quantified by the Loewe Activity score, which is calculated based on the Loewe Additivity model [3]. By using the Loewe score, a drug pair with Loewe score bigger than zero is treated as synergistic, while less than zero is regarded as antagonistic. The higher a Loewe score of combinations, the more effective performances for clinical therapy. To take the noisy data and labels into consideration [1], we choose 10 as our synergy threshold. This means that drug pairs with Loewe score higher than 10 will be labeled as a positive one. While the pairs with score less than 0 will be labeled as negative.

The process of drug features formulation is depicted in Fig. 1 (a). Firstly, the chemical SMILES [4] of drugs were taken from the DrugBank database [5]. Secondly, we utilize RDKit [6] to reformulate the chemical structural formula of a specific drug and convert them into graphs based on the drug's SMILES strings. In the constructed graphs, the atom and chemical bond are represented by node and edge, respectively. The feature vectors of atoms include the information of atom symbol, the number of adjacent atoms, aromaticity, valence, and the number of adjacent hydrogens. The chemical bond feature vectors were encoded according to the information of bond type, the direction of bond, conjugation, aromaticity, ring and owning mol, respectively.

In this case, the gene expression profiles of carcinoma cell lines are derived from Cancer Cell Line

Encyclopedia (CCLE) [7], which is employed to characterize mRNA expression data, genomes and anti-cancer drug dose responses across carcinoma cell lines. We extract the representative genes from the original gene expressions based on a LINCS project [8]. Building upon the Connectivity Map (CMap) data [9], this project provides approximately 1000 well-chosen genes data that covering 80% of the gene information. The process of acquiring gene expression data is depicted in Fig. 1 (b).

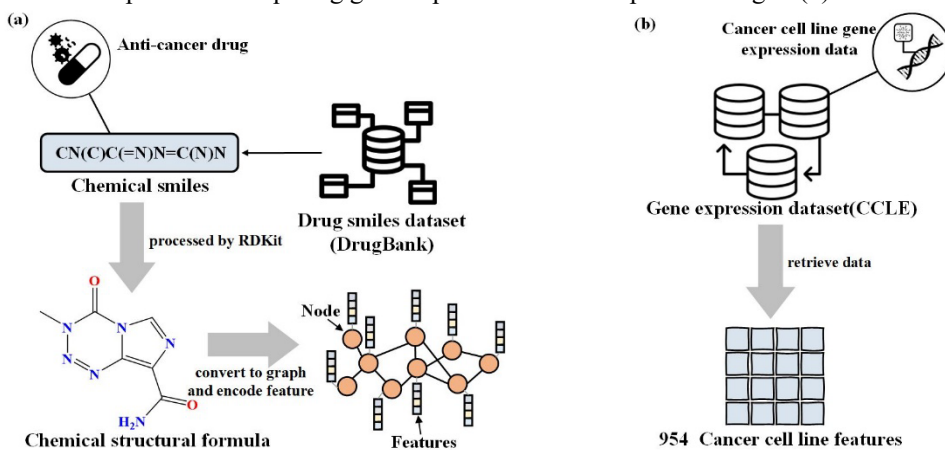

**Fig. 1**. Data collections description. (a) Drug features formulation (b) Cancer cell line feature retrieval

## 2. Cancer cell line feature extraction

We also select 954 numeric features from raw gene expression data as the input of cell line feature reduction block, which is illustrated in Fig. 2. The feature reduction block contains a two layers MLP and a fully connected layer. The computation of cancer cell line feature embeddings can be formulated as follows:

$$\begin{aligned} O_1 &= Dropout\big(\sigma(xW_1 + b_1)\big), \\ O_2 &= Dropout\big(\sigma(O_1 W_2 + b_2)\big), \\ x' &= \sigma(O_2 W_3 + b_3), \end{aligned} \quad (1)$$

where $x$ denotes the input cancer cell line features, $Dropout(\cdot)$ denotes a dropout layer, $\sigma(\cdot)$ denotes an activation function, e.g. ReLU, $W_1, W_2, W_3$ represent the weights, $b_1, b_2, b_3$ are the biases and $x'$ is the output embeddings. After processed by the cell line feature reduction block, the feature embeddings of a specific cell line can be extracted.

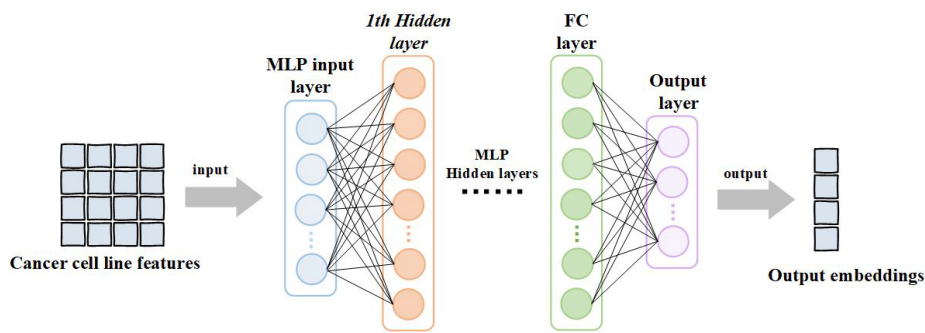

**Fig. 2**. Illustration of cell feature reduction block.

## 3. Chemical feature representation

In EGTSyn, we utilize a molecule to represent a drug. A molecule is comprised of atoms and chemical bonds. In the graph, the atoms are represented by nodes and chemical bonds is the connections between these atoms. The features of each atom are composed of atom symbol, the number of adjacent atoms, aromaticity, valence, and the number of adjacent hydrogens. The features of a chemical bond consist of bond type, the direction of bond, conjugation, aromaticity, ring and owning mol.

In our case, both atoms and chemical bonds of drugs are converted to nodes according to RDKit and served as the inputs of the model. The combination of these two kinds of nodes enables more richer global features to be captured. More details will be demonstrated in the ablation study.

## 4. Workflow of our method

The architecture of EGTSyn is illustrated in Fig. 3. The inputs of the model include two chemical structural SMILES [4] of drug pairs and an untreated cancer cell line expression gene profile. Firstly, the feature graphs of drugs constructed from the input SMILES are fed into our GTDBlock. The GTDBlock utilizes an EGNN and a Transformer-architecture encoder to learn the drug embeddings. Secondly, the gene expression profile of cancer cell line is encoded by the cell line feature reduction block. Thirdly, the obtained drug embeddings pairs and the cell line embeddings are concatenated to form the two drug-cell feature representation vectors, respectively. These two representations are further concatenated together into a single vector and then forward propagated to the following fully connected (FC) layers to form the final prediction label.

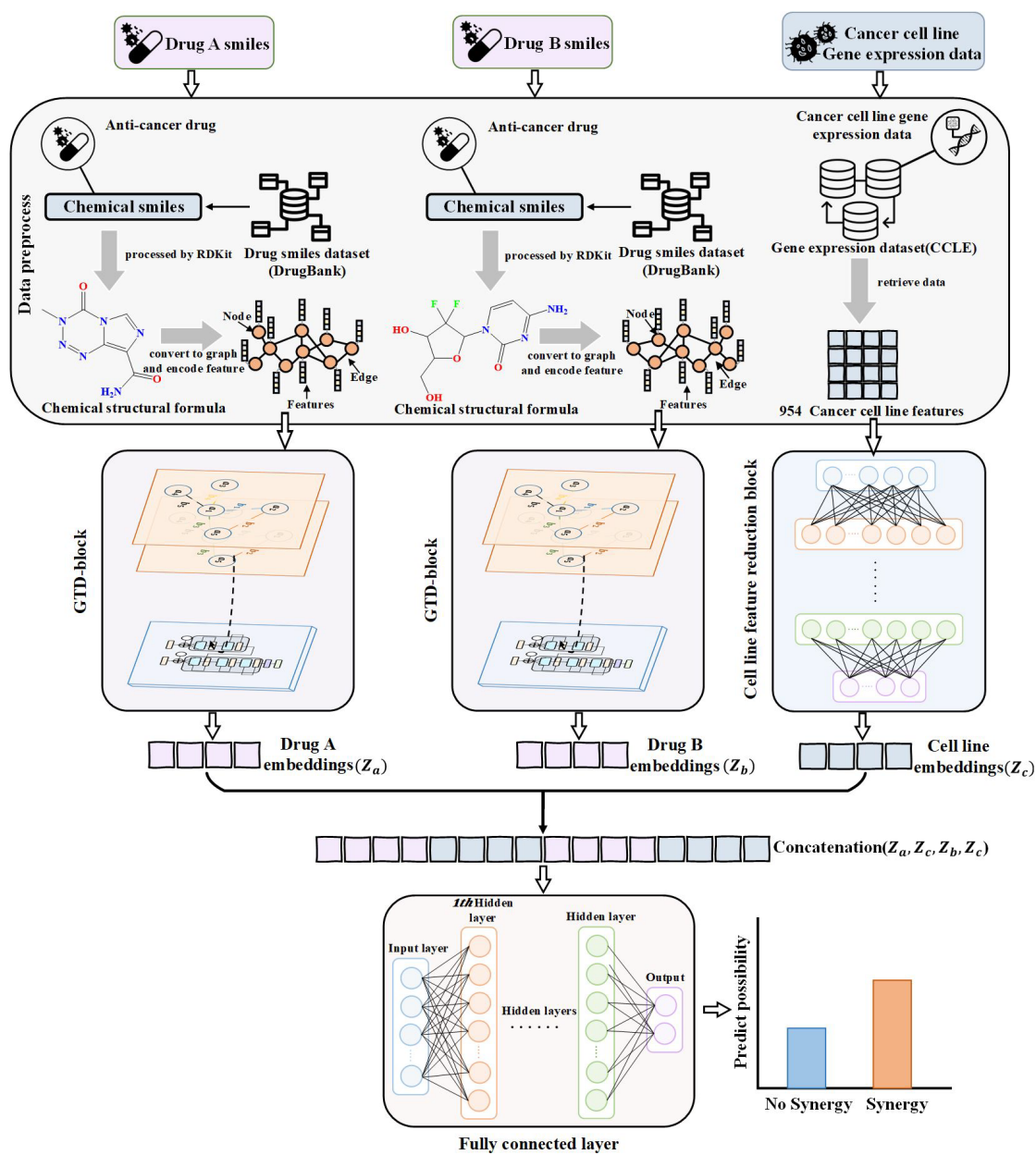

**Fig. 3**. The framework of EGTSyn. The chemical structural SMILES [4] of each drug are pipelined by a GTDblock and then be concatenated with cancer cell line features that calculated by a feature reduction block, respectively. The above two drug-cell feature representation vectors will be further concatenated to serve as input for the final classify layers.

## 5. GTD-block

A GTD-block is a transformer encoder architectural block. The GTD-block takes chemical feature information of drugs as input. Then a EGNN modules is utilized to code the inputs to the intermediate output embeddings. After that, a multi-head attention block followed by a feed forward layer is used to formulate the output intermediate embeddings. The pipeline of GTD-block is depicted in Fig. 4. Details of these blocks will be discussed in the following subsections.

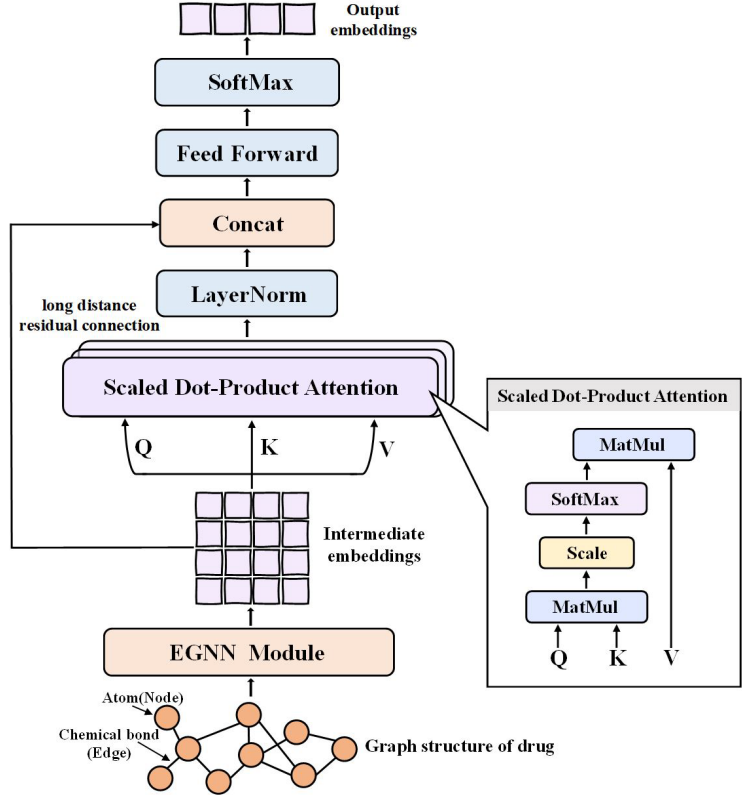

**Fig. 4**. The framework of GTDblock. The converted graphs serve as the input of GTDblock and then be calculated by EGNN module. The intermediate output embeddings will be further duplicated triply as K, Q and V for the following transformer encoder block. Additionally, a long-distance residual connection is also applied to convey low-level structure information.

*5.1. GNN and GCN*

Prescriptively, a graph with node set $V$ and edge set $E$ is denoted as $G = (V, E)$, in which $E \in V \times V$. Specifically, $N_v = |V|$ and $N_e = |E|$ denote the number of nodes and edges. Let $A \in \mathbb{R}^{N_v \times N_v}$ be the adjacency matrix of $G$ and each element $A(i,j)$ denotes the connectivity of node $i$ and node $j$, where $i, j \in V = \{n_1, n_2, \ldots n_{N_v}\}$. Let $X \in \mathbb{R}^{N_v \times d_v}$ be the node feature matrix, where each node is associated with a $d_v$ dimensional feature vector. For a GNN (*Graph Neural Network*), each node updates its representation by aggregating representations of its neighbors layer by layer. The input of a GNN is a node feature matrix $X$ along with an adjacency matrix $A$ and the output is a new node feature matrix $X' \in \mathbb{R}^{N_v \times d}$. The nodes of each layer are computed by the same function:

$$X' = f(X, \psi), \quad (2)$$

where $d$ denotes the output feature dimension, $f$ denotes activation function and $\psi$ is aggregate function.

The design of $f$ and $\psi$ distinguishes the types of GNNs to a large extent. GCN (*Graph Convolutional Network*) [10] is a kind of GNNs and a multi-layer GCN can be propagated as the following rule.

$$H^{(l+1)} = \sigma(\widetilde{D}^{-\frac{1}{2}} \widetilde{A} \widetilde{D}^{-\frac{1}{2}} H^{(l)} W^{(l)}), \quad (3)$$

where $H^{(l)}$ denotes the input matrix of $l$-th layer and $H^{(l+1)}$ is the output of $(l+1)$-th layer. $\widetilde{A} = \widetilde{A} + I_N$ ($I_N$ is identity matrix) is an adjacency matrix with self-connections. $\widetilde{D}_{ii} = \sum_j \widetilde{A}_{ij}$ and $W^l$ is a learnable weight matrix of a specific layer. $\sigma(\cdot)$ denotes an activation function, such as ReLU.

*5.2. EGNN (Edge-based Graph Neural Network)*

EGNN utilizes GCN layers with ReLU activation function to extract the graph features. As illustrated in Fig. 5, two types of graphs, including the atom-based graph and the atom-bond-based graph, were constructed and served as the input of EGNN. For the atom-based graph, the nodes represented atoms and edges indicated chemical bonds, which expressed the structure information of chemicals. As for the designed atom-bond-based graph, the atoms and bonds were both treated as nodes and the relations between them were regarded as edges. In this way, EGNN was capable of capturing the global feature information and structure information of the chemicals as well as the importance of chemical bonds that was neglected by most of previous studies. For example, there will be an atom-bond-based edge between a specific atom and a specific chemical bond in the atom-bond-based graph if the atom was linked to other atoms via the chemical bond.

The two graphs were served as the inputs of EGNN and computed parallelly by the GCN block, respectively. The corresponding outputs will be concatenated as the intermediate output embeddings for the following transformer encoder block. The original GCN targeted at the node-level tasks while the drugs synergistic effect prediction was a graph-level task. We thereby utilized sum, average and max pooling operations to obtain the whole graph feature information from trained node and edge features. Then the aggregated graph features were used to estimate the performances. We found that max pooling operation in the proposed EGTSyn outperformed the other operations. Hence, a global max pooling layer was added after the GCN layer to extract the graph representation.

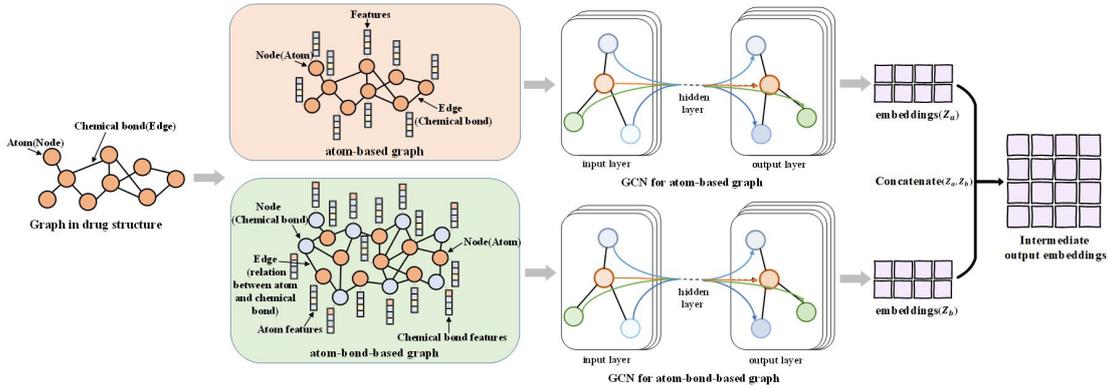

**Fig. 5**. The structure of EGNN. EGNN takes atom-based graph and atom-bond-based graph as inputs. The two graphs are processed by two identical structured GCNs, respectively. The two output embeddings are catenated together to formulate the intermediate embeddings for the following transformer encoder block.

*5.3. Transformer encoder block*

The transformer encoder is built mainly upon the original implementation of Transformer [11], of which the self-attention layer is the basic module. The transformer encoder stacks multiple scale-dot attention layer to formulate multi-head attention. In the GTD-block, a group of queries, keys, values (q, k, v) are taken as the inputs of the self-attention layer. And the q, k and v are the extracted intermediate embeddings of EGNN. The outputs of scale-dot attention can be formulated as follows:

$$Attention(Q, K, V) = softmax(QK^T/\sqrt{d})V, \qquad (4)$$

where $d$ is the dimension of q and k. Assume there are $k$ attention layers in a multi-head self-attention module, then the output matrix is:

$$MultiHead(Q, K, V) = Concat(head_1, \dots, head_k)W^o,$$
$$head_i = Attention(QW_i^Q, KW_i^K, VW_i^V), \qquad (5)$$

where $W_i^Q, W_i^K, W_i^V$ are the projection matrices of head $i$.

It should be noticed that the inputs of a typical self-attention block should be vectorized, such as that in Equation (4). As a matter of fact, the input data of graph is generally not vectorized. Inspired by the previous studies [12, 13], we took the extracted vectors pipelined by EGNN as keys, queries and values of the transformer encoder block. The keys, queries, values will be processed by Scaled Dot-Product Attention [11] with a layer normalization (LN) [14]. Additionally, we also applied a long-distance residual connection structure to concatenate the embeddings processed by EGNN module to convey the initial atom-bond structure information instead of only high-order semantic information being propagated in the network. Then the processed vectors were fed into a feed-forward layer (FFN). In a sense, the residual structure fuses the original structure information between atom-atom and atom-bond as well as the second-level semantic information.

The EGNN module along with a Transformer encoder block together comprise GTDblock (*Graph*

*Transformer for Drugs block*), as shown in Fig. 4.

*6. Predictions*

After being preconditioned, the representations of drugs are fed into the GTDblock and the representations of cancer cell lines are fed into feature reduction block. After pipelined by different modules, the cell line gene expression features will be concatenated to drug A and drug B embeddings. And we can obtain cell line-drug A and cell line-drug B representations. Further, the above two representations are concatenated together to generate the final representation. Finally, the final representation is fed into the classify layer to predict whether synergistic or not.

In EGTSyn, we adopt the cross-entropy as the loss function. It can be formulated as follows:

$$Loss = \min\left(-\sum_{n=1}^{N} \log P_n + \frac{2}{\delta}\|\theta\|\right). \qquad (6)$$

Where $\theta$ refers to all learnable parameters of weight and bias, $N$ is the number of all training samples, $\delta$ is $L2$ norm parameter, while $P_n$ is the predicted possibility of the $n$-th label.

**Experiments and Results**

*1. Hyperparameter settings*

In EGTSyn, the dimension of chemical atom and eigenvectors of chemical bonds were both set as 78, and the dimension of cell line was 945. Considering the vast searching space of the hyper-parameters, grid-search was an available method to tune the parameters of our model and we tuned the hyper-parameters in the experiment of five-folds cross validation on our benchmark dataset. We also conducted a series of experiments for our model with different EGCN layer, EGCN size, GTDblock hidden size, the number of attention heads, as well as the size of cell line feature reduction block MLP and the size of the final linear prediction block. For dropout rate and learning rate, there were multiple values to be evaluated on them. Considering global drug feature, we also applied three graph pool methods to our model and found that the max feature aggregation method behaved better. The best values of these parameters that we tested on our benchmark dataset were demonstrated in boldface.

*2. Performance metrics*

We formulated the drug combination synergy prediction as a classification problem, where the predicted label 1 indicated drug combinations were synergistic while the label 0 was opposite. The measurement metrics covered receiver operator characteristics curve (ROC-AUC), the area under the precision–recall curve (PR-AUC), accuracy (ACC), balanced accuracy (BACC), precision (PREC), True Positive Rate (TPR) and the Cohen's Kappa value (KAPPA).

*3. Compared methods*

We adopted several baseline models to be compared with our proposed method EGTSyn, covering the top-ranking deep learning models and traditional machine learning based models. Four deep learning-based methods were MLP, DeepSynergy, TranSynergy and DeepDDS. Three machine learning methods included Support Vector Machine (SVM), Random Forest (RF) and Adaboost. The proposed method was compared with the baseline methods carrying the best parameter configuration on our benchmark dataset, where 5-fold cross-validation, leave-drugs-out, leave-tissue-out and leave-drug combination-out experiments were evaluated, respectively.

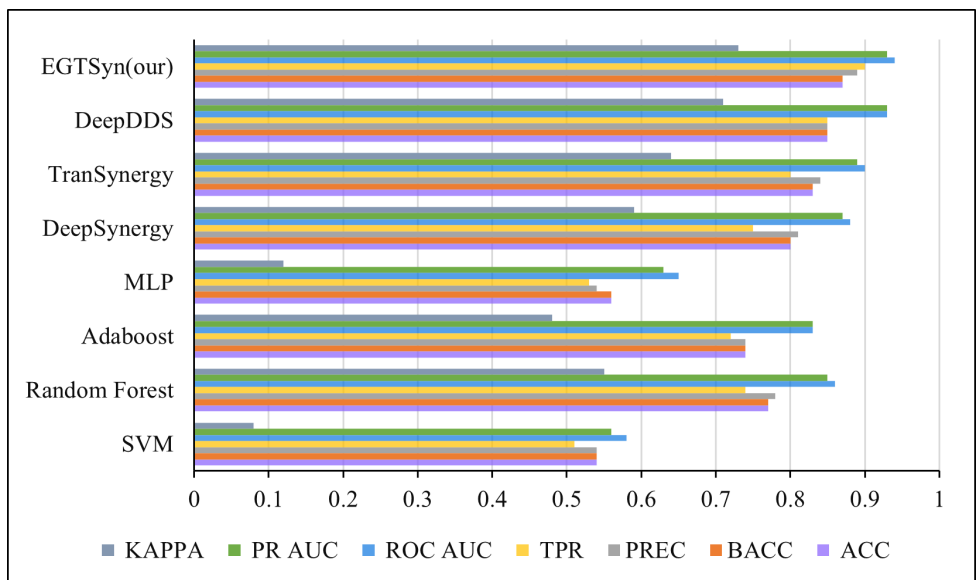

**Fig. 6**. Performance Comparison of EGTSyn with baseline methods upon the random split 5-folds cross validation.

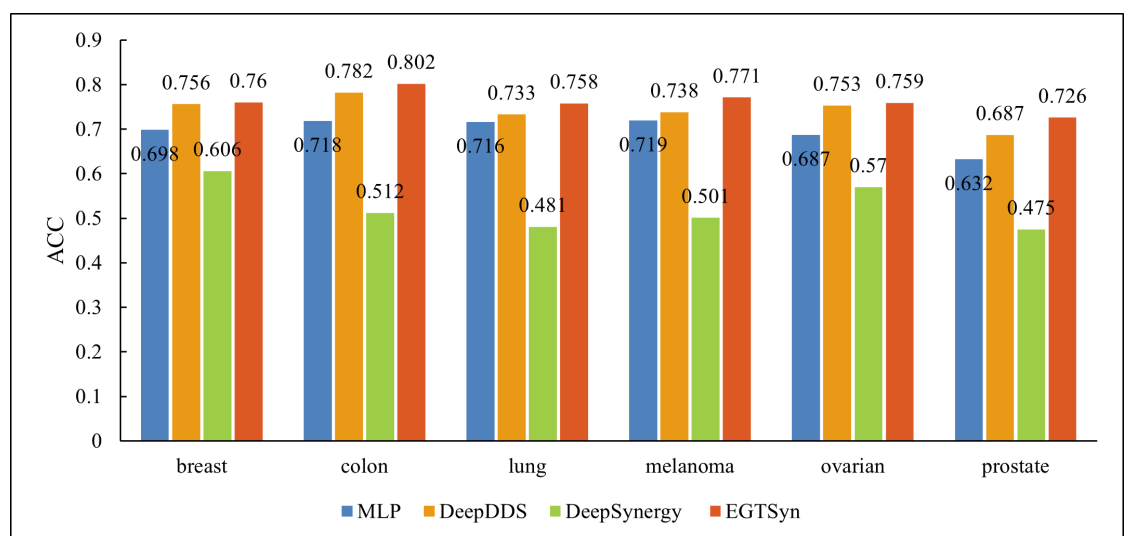

**Fig. 7**. ACC metric of MLP, DeepDDS, DeepSynergy and EGTSyn on six different tissues: breast, colon, lung, melanoma, ovarian, prostate.

## *4. Ablation study*

In this part, we conducted ablation experiments to quantify the contribution of each design in EGTSyn on our benchmark dataset. The results of ablation experiments were included in Table 6. Three variants derived from EGTSyn model were introduced as follows:

**No edge encoding**. We removed the edge encoding block from our model and named it as GTSyn. This means only node embeddings were sent to the graph-based message passing network and then a transformer encoder followed.

**No transformer encoding**. Without considering two-level information extraction of chemicals, we removed the multi-head self-attention mechanism from our model and called it EGSyn. The EGSyn just encoded the atom feature and the atom-bond feature, which would be served as the inputs of the graph-based message passing network respectively, and then their outputs would be concatenated for a pipeline block which consists of some FC layers.

**No both edge and transformer encoding**. To prove the effectiveness of the combination of our proposed two modules, we directly removed the transformer encoder and edge encoder from the EGTSyn with only node embeddings and a graph-based message passing network left for the feature extraction of chemicals. The variant was named GSyn.

As shown in Table 6, our EGTSyn performed the best, and GTSyn as well as EGSyn both second to it. Notably, GSyn performed the worst compared to the other three variants. GTSyn performed slightly better than EGSyn over some metrics, which indicated that the edge encoder module with the information of chemical bonds as well as the self-attention mechanism module in our model both work, and the transformer encoder may contribute more to the result of our method.

Table 6. Results of ablation studies.

|        | ROC AUC   | PR AUC    | ACC       | BACC      | KAPPA     |
|--------|-----------|-----------|-----------|-----------|-----------|
| **EGTSyn** | **0.94±0.05** | **0.93±0.03** | **0.87±0.08** | **0.87±0.08** | **0.73±0.01** |
| GTSyn  | 0.93±0.04 | 0.93±0.07 | 0.86±0.07 | 0.86±0.07 | 0.72±0.01 |
| EGSyn  | 0.93±0.02 | 0.93±0.04 | 0.85±0.04 | 0.86±0.04 | 0.71±0.08 |
| GSyn   | 0.92±0.04 | 0.92±0.05 | 0.85±0.06 | 0.85±0.06 | 0.70±0.01 |

**Discussion and Conclusion**

In this work, a novel EGTSyn model has been proposed for anti-cancer drug combination synergistic effect prediction. EGTSyn take advantages of EGNN module and GTDblock for drug combination features extraction. The EGNN module incorporate the global feature information of the chemicals as well as the importance of chemical bonds that is neglected by most of previous studies. GTDblock combines EGNN with a Transformer encoder block to extract high-order semantic information of drugs. This design enables EGTSyn to capture the global information of chemicals and the corresponding carcinoma cell line gene profiles, which facilitates a more accurate synergetic judgement of drug combinations. The ablation study shows that each design in EGTSyn all contributed to the final prediction. The proposed EGTSyn has been compared with the baseline methods including state-of-the-art models. The results shows that it has superior performance over these competitive baseline methods. In addition, we tested our trained model on the unseen drug pairs and the results verified the superior generalization ability of EGTSyn.

Despite the above advantages, the model still had some limitations. First, EGTSyn doesn't performs very well in the generalization test. This may attribute to the limited number of training samples and the lack of diversity of anti-cancer drugs. Second, the high training cost makes it difficult in handling training datasets expanding. In addition, it is tough to analyze the interpretability of EGTSyn model from the view of physicochemical mechanism limited by the neural network architecture and high complexity of the model.

In conclusion, the proposed EGTSyn is of significant and is quite promising for further wet-lab research. In the future, we will design a light weight model to reduce the high complexity. Also, we will incorporate chemical bond information and high-order semantic information to optimize the EGTSyn model so that it can be trained by a larger dataset with lower cost, and apply it to multi-drug combination synergy prediction tasks as well as more of drug combination therapy fields.